\documentstyle[epsfig]{camera}

\newcommand{\be}{\begin{equation}}
\newcommand{\ee}{\end{equation}}
\newcommand{\bea}{\begin{eqnarray}}
\newcommand{\eea}{\end{eqnarray}}
\begin{document}

%
\title{TOPICS IN CP VIOLATION}

%
\author{Fulvia De Fazio}

%
\organization{Istituto Nazionale di Fisica Nucleare, Sezione di
Bari, Italy}

\maketitle

\begin{center}{\bf Abstract}\end{center}
I briefly review some results and  open questions in the analysis
of CP violation in the B sector.

\section{Introduction}
Although we know since almost forty years that CP is violated, we
still have  a lot to learn about the mechanisms of such a
violation. In particular, we do not know whether the Standard
Model of elementary particle Physics (SM) provides an adequate
description of this phenomenon. CP violation is one of the
necessary conditions to explain matter/antimatter asymmetry
starting from initially symmetric conditions in the early
Universe. Moreover, baryogenesis requires sources of CP violation
beyond SM and actually almost all new Physics scenarios provide
such new sources. Therefore,  CP violation is an efficient testing
ground for the SM. CP violation was observed first in  neutral
kaon decays through the detection of the mode $K_L \to \pi \pi$
\cite{Christenson:fg}. One defines $\epsilon=\displaystyle{A(K_L
\to \pi \pi (I=0)) \over A(K_S \to \pi \pi (I=0))}$, i.e. the
ratio of a CP suppressed decay to a CP allowed one. The
experimental measure: $|\epsilon|_{exp}=(2.271 \pm 0.017) \times
10^{-3}$ \cite{Hagiwara:fs} probes the mechanism of {\it indirect}
CP violation, due to the fact that the particles taking part into
weak processes are not CP eigenstates, although the amplitudes
themselves do not violate CP.
 The kaon phenomenology
provided us also with the first measure of {\it direct} CP
violation, i.e. directly at the decay amplitude level, through the
measure of Re$ \left( \displaystyle{\epsilon^\prime / \epsilon}
\right)=(1.8 \pm 0.4) \times 10^{-3} $ \cite{Hagiwara:fs}. From
the theoretical side, many uncertainties affect the calculation of
this parameter, essentially linked to the matrix elements of the
effective  weak hamiltonian $H_{eff}(\Delta S=1)$ between the kaon
and a two pion state.

Within the SM, the only source of CP violation is  the
Cabibbo-Kobayashi-Maskawa matrix $V_{CKM}$ which describes the
mixing of down-type quarks in charged current interactions.
 In the most general pattern, $V_{CKM}$
depends on four parameters, one of which is a complex phase. In
the Wolfenstein parametrization they  are $\lambda, A, \rho$ and
the complex phase  $ \eta$.

\begin{figure}[ht]
\begin{center}
\vspace*{-3.6cm} \mbox{\epsfig{file=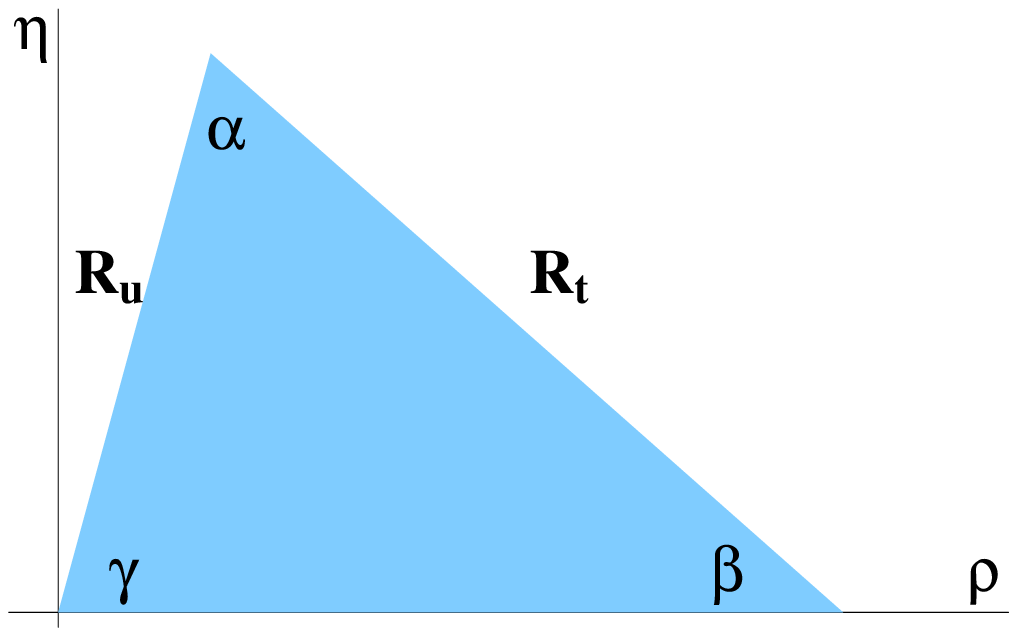, width=5cm}}
\vspace*{-0.5cm} \
\end{center}
\caption{\baselineskip 15pt Unitarity Triangle} \label{ut}
\end{figure}

The SM requires $V_{CKM}$ to be unitary. The  unitarity relation
$V_{ud}V_{ub}^*+V_{cd}V_{cb}^*+V_{td}V_{tb}^*=0$ can be
represented as a triangle in the $(\bar \rho, \bar \eta)$
plane,where ${\bar \rho}=\rho(1-\lambda^2/2)$, ${\bar
\eta}=\eta(1-\lambda^2/2)$, as shown in Fig.
\ref{ut}.\put(-145,-15){-} \put(-10.5,-93){-}
The angles are linked to the phases of the CKM elements  through
the relations: \be \alpha={\rm Arg} \left( -{V_{td} V_{tb}^* \over
V_{ud} V_{ub}^* } \right) \,\,\, \beta={\rm Arg} \left( -{V_{cd}
V_{cb}^* \over V_{td} V_{tb}^*  } \right)\,\,\, \gamma={\rm Arg}
\left( -{V_{ud} V_{ub}^* \over V_{cd} V_{cb}^*  } \right) \,\,.
\label{angles} \ee Many efforts are devoted to  determine
$\alpha,\, \beta, \, \gamma$ and  constrain the upper vertex of
the triangle. The measure of $\epsilon$ constrains such a  vertex
in a region of the $(\bar \rho,\bar \eta)$ plane determined by two
hyperbolae. Analogously,  semileptonic B decays give information
about the length  $R_u$ of the left side, providing again a region
for the apex. More constraints can be obtained from the B
phenomenology. Many  reviews are available on this subject
\cite{Ball:2000ba}, here we shall briefly recall some results and
open questions in the B sector.

\section{CP violation in the system of neutral $B$ mesons}

Neutral B mesons mix with each other, the mixing being described
by a box diagram dominated by the intermediate top exchange
contribution, with Arg(box)=2 Arg[$V_{td}V_{tb}^*$]=2$\delta^m=2
\beta$ (for B$_s$ system: $\delta^m_s$= Arg[$V_{ts}V_{tb}^*$]).
Due to oscillations, the system evolves and time-dependent
CP-asymmetries can be considered. It is particularly convenient to
consider decays into a CP eigenstate $f_{CP}$ in which both $B^0$
and ${\overline B}^0$ can decay. The asymmetry reads: \bea
a_{f_{CP}}&=&{  \Gamma(B^0(t) \to f_{CP})- \Gamma({\bar B}^0(t)
\to f_{CP}) \over \Gamma(B^0(t) \to
f_{CP})+ \Gamma({\bar B}^0(t) \to f_{CP})  }= \nonumber \\
&=& {(1-|\lambda|^2) {\mathrm cos}(\Delta m t)-2 {\mathrm Im}
\lambda {\mathrm sin} (\Delta m t)
\eta_f \over 1+|\lambda|^2  }= \label{cpasym}\\
&=&C_{f_{CP}} {\mathrm cos} (\Delta m t)-S_{f_{CP}} {\mathrm sin}
(\Delta m t) \nonumber
 \eea \noindent
where $\lambda=\displaystyle{e^{-2i \delta^m} {\bar A} / A}$ and
${\cal A}(B^0 \to f_{CP})=A$, ${\cal A}( \bar B^0 \to f_{CP})=
\bar A$; $\eta_f$ is the CP-parity of the state $f_{CP}$. In
general, one may write: $A=\sum_i A_i e^{i \delta_i} e^{i
\phi_i}$, $\bar A=\sum_i A_i e^{i \delta_i} e^{-i \phi_i}$, the
$\delta_i$ being strong phases, the $\phi_i$ weak ones. The term
$C_{f_{CP}}$ probes direct CP violation, being zero if
$|\lambda|=1$, i.e. if $|\bar A|=|A|$, while the term $S_{f_{CP}}$
probes mixing-induced CP violation.

Special cases occur when   $\forall i: \phi_i=\phi$, so that
$|\bar A|=|A|$,  $|\lambda|=1$ and $a_{f_{CP}}=-{\mathrm Im}
\lambda {\mathrm sin} (\Delta m t) \eta_f$, with  ${\mathrm Im}
\lambda= -{\mathrm sin}[2(\delta^m +\phi)]$. Before considering
which  modes are suitable to extract the angles of the unitarity
triangle, it should be recalled that also the measure of the
oscillation parameters $\Delta m_q$, $\Delta m_s$ provide
constraints in the $(\bar \rho, \bar \eta)$ plane. The mass
difference in the B (B$_s$) system can be obtained from the box
diagram with virtual top exchange, so that $\Delta m_d \propto
|V_{td}|^2$, $\Delta m_s \propto |V_{ts}|^2$. From the world
average: $\Delta m_d=0.502 \pm 0.006 \,\,ps^{-1}$ and the bound:
$\Delta m_s >14.4 \,\, ps^{-1}$, other regions can be selected,
which are circles centered at $(1,0)$.

\section{Strategies to extract $\alpha$, $\beta$, $\gamma$}
Among the angles of the triangle, $ \beta$ has been the first one
which has been experimentally measured through the observation of
the time dependent CP asymmetry in $B_d^0 \to J/\psi K_S$. Within
the SM, this process is induced at quark level by the transition
${\bar b} \to {\bar c} c {\bar s}$, proceeding through tree and
penguin diagrams. The penguin with up-quark exchange is subleading
in the Wolfenstein parameter $\lambda$, so that, neglecting its
contribution, the amplitude is dominated by a single weak phase,
i.e. that of $V_{cb}V_{cs}^*$, which is real to a very good
approximation. This means that the time dependent asymmetry
provides just sin$(2 \beta)$, and hence $2\beta$ up to a twofold
ambiguity. The world average of the experimental determinations is
dominated by the most recent Belle and BABAR data \cite{beta} and
reads: sin$(2\beta)=0.736 \pm 0.049$, giving $2\beta=(47 \pm 4)^o$
or $2\beta=(133 \pm 4)^o$. The first solution is in good agreement
with the indirect determination obtained through the CKM fits
\cite{Battaglia:2003in}, although one cannot exclude {\it a
priori} the second one, which would signal new physics effects.
The resolution of this discrete ambiguity has been subject of
several works \cite{2b}. A possibility is to consider the modes
$B^0 ({\bar B}^0)\to D^+ D^- \pi^0 (K_S)$,  assuming that they
proceed through  intermediate S-wave and P-wave charmed and beauty
resonances \cite{charles,Colangelo:1999ny}. The variables
$s_+=(p_{D^+}+p_\pi)^2$ and $s_-=(p_{D^-}+p_\pi)^2$ can be
introduced, in terms of which the time dependent amplitudes read
as \bea |{\cal A}(B^0(t) \to D^+ D^- \pi^0)|^2&=&{e^{-\Gamma t}
\over 2}[G_0(s_+,s_-)+G_c(s_+,s_-){\mathrm cos}(\Delta mt)
\nonumber
\\ &-& G_s(s_+,s_-){\mathrm sin}(\Delta
mt)] \nonumber \\
|{\cal A}({\bar B}^0(t) \to D^+ D^- \pi^0)|^2&=&{e^{-\Gamma t}
\over 2}[G_0(s_-,s_+)-G_c(s_-,s_+){\mathrm cos}(\Delta
mt)\nonumber
\\
&+& G_s(s_-,s_+){\mathrm sin}(\Delta mt)] \eea where \bea
G_0(s_+,s_-)&=&
|A(s_+,s_-)|^2+ | {\bar A}(s_+,s_-)|^2 \nonumber \\
G_c(s_+,s_-)&=& |A(s_+,s_-)|^2- | {\bar A}(s_+,s_-)|^2 \nonumber
\\
G_s(s_+,s_-)&=& -2 {\mathrm sin}(2 \beta){\mathrm
Re}[A^*(s_+,s_-)| {\bar A}(s_+,s_-)]\nonumber \\ &+& 2 {\mathrm
cos}(2 \beta){\mathrm Im}[A^*(s_+,s_-)| {\bar
A}(s_+,s_-)]\label{gi} \eea and $A(s_+,s_-)={\cal A}( B^0 \to D^+
D^- \pi^0)$, ${\bar A}(s_+,s_-)={\cal A}({\bar B}^0 \to D^+ D^-
\pi^0)$. From (\ref{gi}) one can see that it is possible to access
to ${\mathrm cos}(2 \beta)$. The estimated branching ratios
\cite{Colangelo:1999ny} also make their experimental detection
rather promising.

An interesting test can be carried out considering the mode $B_d
\to \phi K_S$  since its CP asymmetry should coincide, within the
SM, with that of $B_d \to J/\psi K_S$. Recent results give
\cite{Abe:2003yt}: \bea C_{\phi K_S}^{BABAR}=-0.80 \pm 0.38 \pm
0.12 \hskip
1 cm &&  S_{\phi K_S}^{BABAR}=-0.18 \pm 0.51 \pm 0.07 \nonumber \\
C_{\phi K_S}^{BELLE}=0.15 \pm 0.29 \pm 0.07 \hskip 1 cm && S_{\phi
K_S}^{BELLE}=0.96 \pm 0.50^{+0.11}_{-0.09} \label{phik} \,.\eea
Since for $B_d \to J/\psi K_S$  $C_{\psi K_S}=0$, $S_{\psi
K_S}=-{\mathrm sin}(2\beta)$, the  results (\ref{phik}) might
represent hints of new physics, although the large uncertainties
 prevent from drawing any conclusion. Since the
 determination of sin$(2\beta)$ from
$B_d \to J/\psi K_S$ and the indirect one from the fits agree with
each other, eventual new physics would affect the transition
${\bar b} \to {\bar s}s {\bar s}$ more likely than ${\bar b} \to
{\bar c}c {\bar s}$ and hence signals should be detected also in
other modes, such as $B_d \to \eta^\prime K_S$, for which
preliminary results agree with those from $B_d \to J/\psi K_S$.

Let us now consider the status of the determination of $\alpha$.
 Within the SM, if penguin contributions could be
neglected, it would be possible to get $\alpha$ from $B_d \to
\pi^+ \pi^-$. In this case, one would have $C_{\pi \pi}=0$ and
$S_{\pi \pi}={\mathrm sin}(2 \beta+2 \gamma)=-{\mathrm sin}(2
\alpha)$. However, penguins are not negligible and should be taken
into account. The B factories provide us with the results
\cite{Abe:2003ja,Aubert:2002jb}: \bea C_{\pi \pi}^{BELLE}=-0.77
\pm 0.27 \pm 0.08 \hskip
0.6 cm &&  S_{\pi \pi}^{BELLE}=-1.23 \pm 0.41 ^{+0.08}_{-0.07} \nonumber \\
C_{\pi \pi}^{BABAR}=-0.30 \pm 0.25 \pm 0.04 \hskip 0.6 cm &&
S_{\pi \pi}^{BABAR}=0.02 \pm 0.34 \pm 0.03 \label{pipi} \,.\eea In
order to get rid of the penguin contribution, a fit procedure was
adopted in \cite{Abe:2003ja}, comparing the experimental results
to theory; four parameters are involved: $\beta$ which is taken
from the direct measure of sin$(2 \beta)$; the tree to penguin
amplitude ratio $|P|/|T|$, varied according to theoretical
estimates; $\alpha$ and the strong phase difference between $T$
and $P$. The result  $78^o \le \alpha \le 152^o$ is consistent
with the range determined indirectly from the CKM fits.

The determination of $\gamma$ is  affected by  theoretical
uncertainties. We refer to \cite{Ball:2000ba} for a list of modes
suitable to extract this phase. In order to consider strategies
 relevant for future machines, we discuss instead
a proposal to obtain $\gamma$ from B$_c$ decays. B$_c^-$ is the
lowest lying $b{\bar c}$ meson, discovered by CDF Collaboration in
1998 \cite{cdf}. A large number of such particles will be produced
at future hadron colliders and hence it is worthy to consider the
possibility of studying CP violation through its decays. For this
purpose, Fleischer and Wyler \cite{Fleischer:2000pp} considered
the six decay modes: $B_c^\pm \to D_s^\pm \{ D^0, {\bar D}^0,
D_+^0\}$, where   $|D_+^0>=(|D^0>+|{\bar D}^0>)/\sqrt{2}$ is the
CP-even eigenstate. The following relations among the decay
amplitudes hold: \bea \sqrt{2} {\cal A}(B_c^+ \to D_s^+ D_+^0) &=&
{\cal A}(B_c^+ \to D_s^+ D^0)+{\cal A}(B_c^+ \to D_s^+ {\bar D}^0)
\nonumber \\
\sqrt{2} {\cal A}(B_c^- \to D_s^- D_+^0) &=& {\cal A}(B_c^- \to
D_s^- D^0)+{\cal A}(B_c^- \to D_s^- {\bar D}^0) \label{bc} \,.\eea
The amplitudes  in (\ref{bc}) are roughly of the same order, since
the colour suppressed mode $B_c^+ \to D_s^+ {\bar D}^0$ is
enhanced by $V_{cb}$, while the colour allowed $B_c^+ \to D_s^+
D^0$ is suppressed by $V_{ub}$. This is the main difference with
respect to the analogous $B^\pm \to K^\pm D$ modes, where colour
suppressed amplitudes are proportional to $V_{ub}$. Furthermore,
the only weak phase involved is $\gamma$, so that the
 relations in (\ref{bc}) can be represented as two triangles
with a common basis (${\cal A}(B_c^+ \to D_s^+ {\bar D}^0)={\cal
A}(B_c^- \to D_s^- D^0)$), but mismatched by $2 \gamma$.

This method presents many advantages: the sides of the triangles
have similar sizes and only tree diagrams are involved, contrarily
to $B \to K \pi$, also considered to extract $\gamma$, where
penguins are important \cite{Grossman:2003qi}. Therefore,
comparing
 the results of the two methods would be a probe
of new physics. Finally,  $B_c$ decays considered here are likely
to be observed \cite{Colangelo:1999zn} at LHC, where a large
number (${\cal O}(10^{10}))$ of $B_c$ per year are expected to be
produced.

Among the  strategies to over-constrain the unitarity triangle
parameters, an important role will be played by B$_s$ physics.
Although B$_s$ is non currently accessible at B factories, it will
be copiously produced at hadron colliders. At present, the  upper
bound on $\Delta m_s$ already constrains $R_t$, implying $\gamma
\le 90^o$. However, from B$_s$ decays much more information could
be gained, as  from the mode $B_s \to J/\psi \phi$. In complete
analogy with $B_d  \to J/\psi K_S$, this decay will give access to
the phase of B$_s$ mixing.

\section{Conclusions}
\begin{figure}[ht]
\begin{center}
\vspace*{-0.3cm} \mbox{\epsfig{file=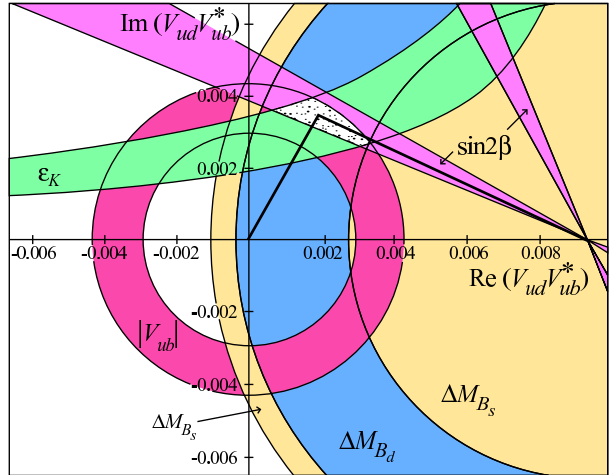, width=8cm}}
\end{center}
\caption{\baselineskip 15pt Constraints to the Unitarity Triangle
from PDG [2].} \label{pdgut}
\end{figure}

Fig. \ref{pdgut} displays several constraints in the $(\bar \rho,
\bar \eta)$ plane as reported by PDG, edition 2003
\cite{Hagiwara:fs}. The resulting scenario shows no
inconsistencies; in particular, the comparison between the direct
determination of sin$(2 \beta)$ and the indirect result of the CKM
fits suggests that at present the SM describes coherently CP
violation. Hence, one should look for small effects to reveal new
physics, for example in $b \to s$ penguins or in $\Delta m_s$. New
observables may be studied to obtain further constraints and a
reduction of theoretical uncertainties affecting the various
predictions should be pursued, while waiting for more results in
the LHC era.

\begin{thebibliography}{}
%
\bibitem{Christenson:fg}
J.~H.~Christenson {\it et al.},
Phys.\ Rev.\ Lett.\  {\bf 13} (1964) 138.
%
\bibitem{Hagiwara:fs}
K.~Hagiwara {\it et al.},
Phys.\ Rev.\ D {\bf 66} (2002) 010001.
%
\bibitem{Ball:2000ba}
See for example: P.~Ball {\it et al.}, B decays at the LHC,
hep-ph/0003238.
%
\bibitem{beta}
B.~Aubert {\it et al.},
Phys.\ Rev.\ Lett.\  {\bf 89}, 201802 (2002);
K.~Abe {\it et al.},
arXiv:hep-ex/0308036.
%
\bibitem{Battaglia:2003in}
M.~Battaglia {\it et al.},
arXiv:hep-ph/0304132.
%
\bibitem{2b}
Y.~I.~Azimov {\it et al.},
Z.\ Phys.\ A {\bf 356}, 437 (1997;
Y.~Grossman and H.~R.~Quinn,
Phys.\ Rev.\ D {\bf 56}, 7259 (1997);
B.~Kayser and D.~London,
Phys.\ Rev.\ D {\bf 61}, 116012 (2000);
H.~R.~Quinn  {\it et al.},
Phys.\ Rev.\ Lett.\  {\bf 85}, 5284 (2000);
A.~S.~Dighe {\it et al.},
Phys.\ Lett.\ B {\bf 433}, 147 (1998);
I.~Dunietz, R.~Fleischer and U.~Nierste,
Phys.\ Rev.\ D {\bf 63}, 114015 (2001).
%
\bibitem{charles}
J.~Charles {\it et al.},
Phys.\ Lett.\ B {\bf 425} (1998) 375 [Erratum-ibid.\ B {\bf 433}
(1998) 441].
%
\bibitem{Colangelo:1999ny}
P.~Colangelo {\it et al.},
Phys.\ Rev.\ D {\bf 60} (1999) 033002.
%
\bibitem{Abe:2003yt}
K.~Abe {\it et al.},
arXiv:hep-ex/0308035;
H. Yamamoto, talk at the International Europhysics Conference on
High-Energy Physics (HEP 2003), Aachen, Germany, 17-23 Jul 2003.
%
\bibitem{Abe:2003ja}
K.~Abe {\it et al.},
Phys.\ Rev.\ D {\bf 68}, 012001 (2003).
%
\bibitem{Aubert:2002jb}
B.~Aubert {\it et al.},
Phys.\ Rev.\ Lett.\  {\bf 89} (2002) 281802.
%
\bibitem{cdf}
F.~Abe {\it et al.},
Phys.\ Rev.\ D {\bf 58}, 112004 (1998);
Phys.\ Rev.\ Lett.\  {\bf 81}, 2432 (1998).
%
\bibitem{Fleischer:2000pp}
R.~Fleischer and D.~Wyler,
Phys.\ Rev.\ D {\bf 62} (2000) 057503.
%
\bibitem{Grossman:2003qi}
For a discussion see: Y.~Grossman,
arXiv:hep-ph/0310229.
%
\bibitem{Colangelo:1999zn}
P.~Colangelo and F.~De Fazio,
Phys.\ Rev.\ D {\bf 61} (2000) 034012.


\end{thebibliography}
\end{document}